\documentclass[amsmath,amssymbs,aps,twocolumn]{revtex4-2}

\usepackage{graphicx}
\usepackage{amsmath,amsfonts,amssymb}
\usepackage{epsfig}
\usepackage{wrapfig}
\usepackage{bbm}
\usepackage[usenames]{color}
\usepackage{array}
\usepackage{times}
\usepackage{float}

\begin{document}

\preprint{APS/123-QED}

\title{Topologically Quantized Soliton-Like Pumping using Synthetic Nonlinearity}
\author{Ankitkumar Maisuriya}
\author{Siddhi Mali}
\author{Sunil Mittal}
\email{Contact author: s.mittal@northeastern.edu}
\affiliation{Department of Electrical and Computer Engineering, Northeastern University, Boston, MA 02115, USA}
\affiliation{Institute for NanoSystems Innovation, Northeastern University, Boston, MA 02115, USA}



\begin{abstract}
The interplay between nonlinear and topological physics has led to intriguing emergent phenomena, such as quantized and fractionally quantized Thouless pumping of solitons dictated by the topological invariants of the underlying band structure. Unlike linear Thouless pumping, which requires excitation of a Wannier function of a uniformly filled band, quantized soliton pumping is observed even with localized excitations that do not represent Wannier functions. Here, we show that similar soliton-like quantized pumping can be observed in Aubry–André–Harper model by introducing a synthetic nonlinearity in the form of a cutoff on the coupling strengths between lattice sites. More importantly, we reveal that the localized excitations driving quantized soliton pumping are precisely the Wannier functions of the uniformly filled bands of the effectively nonlinear lattice, thus restoring consistency with linear Thouless pumping. We extend this approach to multi-band systems and show that the nonlinearity introduces a degeneracy between bands, subsequently leading to the observation of fractionally quantized pumping. Our approach of introducing a synthetic nonlinearity is general and could be extended to reveal soliton dynamics in other nonlinear topological systems. 
\end{abstract}
                
\maketitle

Quantized pumping in periodically driven topological systems is a manifestation of quantized transport, where the particle displacement per pump cycle is determined by the topological invariant of the underlying band structure \cite{Thouless1982, Thouless1983, Niu1984, Citro2023}. A hallmark example is the Thouless pump in the Aubry–André–Harper (AAH) model, which connects quantized pumping in driven one-dimensional (1D) systems to the two-dimensional static Harper–Hofstadter model through dimensional reduction \cite{Aubry1980, Harper1955, Kraus2012, Citro2023}. The recent emergence of synthetic topological systems has enabled the observation of topological pumping across a range of platforms, including photonic systems of coupled waveguides and resonators \cite{Kraus2012, Ke2016, Zilberberg2018, Sridhar2024, Haldane2008, Lu2014, Khanikaev2017, Ozawa2019, Mehrabad2023}, acoustic \cite{Xu2025, Xue2022} and mechanical systems \cite{Huber2016, Grinberg2020}, lattices of ultracold atoms \cite{Lohse2016, Nakajima2016, Lohse2018}, and superconducting circuits \cite{Liu2025}. 

The presence of nonlinearities in such synthetic systems has further enabled the remarkable observation of topological pumping of self-localized wavepackets \cite{Jurgensen2021,Jurgensen2022, Mostaan2022, Ke2017}, called solitons, that balance nonlinear effects against diffraction to preserve their shape during evolution \cite{Eisenberg1998, Christodoulides1988, Fleischer2003}. More interestingly, quantized Thouless pumping in the linear AAH model necessitates the excitation of a Wannier function of a filled band such that the center of mass of the wavepacket is displaced by an integer number of unit cells equal to the Chern number of the corresponding band \cite{Thouless1983, Ke2016, Jurgensen2021}. Nevertheless, solitons exhibit quantized pumping even when the excitation, localized to a single or a few lattice sites, does not correspond to the Wannier function of the filled band of the linear lattice from which they bifurcate \cite{Jurgensen2021, Jurgensen2022, Mostaan2022}. Subsequently, this phenomenon has also enabled the observation of fractionally quantized pumping where solitons bifurcate from multiple bands, and are therefore, displaced by only a fraction of the unit cell in one period \cite{Jurgensen2023, Jurgensen2025}.

\begin{figure*}[ht!]
    \centering
    \includegraphics[width=1\textwidth]{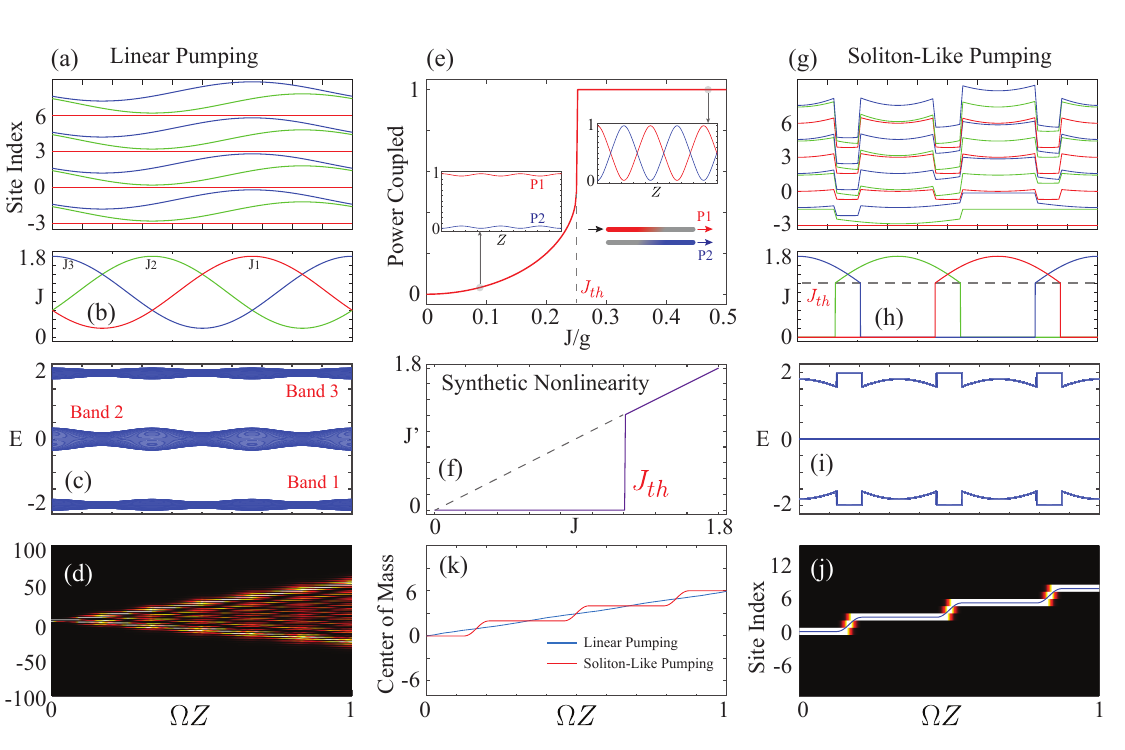}
    \caption{\textbf{a.} Schematic of a 1D coupled waveguide array implementing the AAH model with periodicity $N = 3$. \textbf{b.} Modulated coupling strengths as a function of the propagation distance $z$. The y-axis represents the positions of different waveguides of the unit cell; colors distinguish the three waveguides in a unit cell. \textbf{c.} Band structure of the linear AAH lattice, with periodic boundary condition (PBC), as a function of the propagation distance $z$. \textbf{d.} Spatial evolution and quantized pumping in the linear AAH lattice, for excitation of a Wannier function corresponding to the second bulk band. \textbf{e.} Coupled power in a pair of nonlinear waveguides, showing a sharp transition with increasing nonlinear strength $g$. \textbf{f.} The nonlinearity-induced transition in \textbf{e} is mimicked by introducing a synthetic nonlinearity, in the form of a threshold in coupling strength $J$. \textbf{g-i} Schematic of the waveguide array, modulation of coupling strengths, and band structure of the modified, effectively nonlinear lattice. \textbf{j} Soliton-like pumping in the nonlinear lattice by two unit cells, corresponding to the second band with Chern number 2. \textbf{k}. Comparison of linear and soliton-like pumping, both showing the same displacement over a drive period.}
    \label{Fig:1}
\end{figure*}

Here, we show that similar soliton-like quantized Thouless pumping can be realized in the corresponding linear systems by simply introducing a synthetic nonlinearity in the lattice such that coupling strengths below an appropriate threshold are set to zero (Fig.\ref{Fig:1}). More importantly, we show that although the single- or few-site excitation for soliton-like quantized pumping does not correspond to the Wannier function of any filled band of the original linear lattice, it does represent the Wannier function of the thresholded, effectively nonlinear lattice. This observation restores the conventional understanding that quantized topological pumping requires excitation of a Wannier function of a uniformly filled band. We extend this formalism to fractionally quantized soliton-like pumping and show that the emergence of fractional quantization is due to the nonlinearity-induced degeneracy between two bands of different Chern numbers. We also extend this formalism to pumping in 2D systems, which, through dimensional reduction, are connected to the 4D quantum Hall effect \cite{Kraus2013, Zilberberg2018, Lohse2018}. Although we specifically use the AAH model to demonstrate soliton-like pumping, our formalism of introducing synthetic nonlinearity in the form of thresholded coupling strengths could be applied to understand the existence and evolution of topological solitons in other nonlinear systems, including those with non-Hermitian and higher-order topological phenomena \cite{Maczewsky2020, Mukherjee2020, Xia2021, Fedorova2020, Fu2022, Arkhipova2023, You2025, Smirnova2020, Szameit2024, Benalcazar2017, Schindler2018, Mittal2021, Hashemi2024}.  


We consider a 1D lattice implementing the off-diagonal nonlinear AAH model. The spatial/temporal
evolution of the wavefunction $\psi_{n}$, at a lattice site $n$, is described by the equation
\begin{equation} \label{NLEvol}
i\frac{d\psi_{n}}{dz} = -J_{n} \psi_{n+1} - J_{n-1} \psi_{n-1} - g \left|\psi_{n}\right|^2 \psi_{n}.
\end{equation} 
Here,  $g$ is the strength of nonlinearity. $J_{n}$ is the coupling strength between the lattice sites $n$ and $n+1$, and it oscillates sinusoidally as a function of site index and evolution length $z$ (or, equivalently, time $t$) as
\begin{equation}
 J_{n}\left(z\right) = J_{0} + \delta ~cos\left(\frac{2\pi}{N} n + 2\pi \Omega z + \phi_{0} \right),
\end{equation}
where $J_{0}$ is the mean coupling strength, $\delta$ is the modulation amplitude. $\Omega = \frac{1}{\Lambda}$ such that $\Lambda$ is the drive period and $\phi_{0}$ is a phase offset. $N$ is the number of lattice sites in a unit cell such that $\Phi = \frac{2\pi}{N}$ is the magnetic flux for the equivalent Harper-Hofstadter model. This nonlinear lattice can be realized, for example, using an array of coupled optical waveguides \cite{Jurgensen2021, Jurgensen2023}. In the absence of nonlinearity ($g = 0$), the spatial (temporal) evolution of an input excitation exhibits diffraction (dispersion) in the lattice, while its center-of-mass is displaced by an integer number of unit cells (Fig.\ref{Fig:1}d). Nevertheless, in the presence of nonlinearity, the input excitation can generate solitons that propagate through the lattice without any diffraction, but still exhibit quantized and fractionally quantized Thouless pumping \cite{Jurgensen2021, Jurgensen2022, Mostaan2022, Jurgensen2023}.

\begin{figure*}[ht!]
    \centering
    \includegraphics[width=\textwidth]{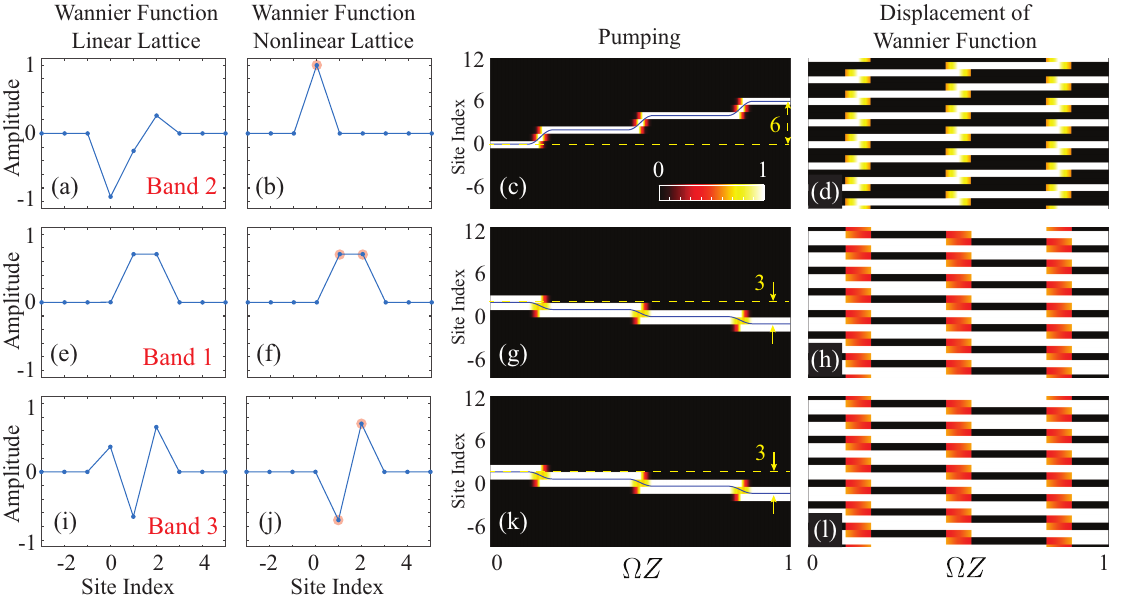}
    \caption{Wannier functions corresponding to Band 2, for \textbf{a}, linear lattice, and \textbf{b}, effectively nonlinear lattice. The shaded red circles show the excitation wavefunction. \textbf{c.} Quantized pumping for Band 2 by 2 unit cells per period. The solid blue line shows the displacement of the CoM. \textbf{d.} Instantaneous Wannier function intensity in the lattice, showing displacement by Chern number $+2$. \textbf{e-h.} Corresponding results for Band 1, showing pumping by one unit cell, but in the opposite direction. \textbf{i-l} Results for Band 3, also showing pumping by one unit cell.}
    \label{Fig:2}
\end{figure*}

We introduce a synthetic nonlinearity in the form of modified coupling strengths $J^{'}_{n}$, such that $J^{'}_{n}\left(z\right) = J_{n}\left(z\right)$ if $J_{n} \geq J_{Th}$, but $J^{'}_{n}\left(z\right) = 0$ if $J_{n} < J_{Th}$ (Fig.\ref{Fig:1}f). Here $J_{Th}$ is an appropriately chosen threshold coupling strength that corresponds to the strength of nonlinearity. As such, the spatial/temporal evolution of the wavefunction in the linear lattice ($g = 0$), but with an effective synthetic nonlinearity introduced using thresholded coupling strengths, is described as
\begin{eqnarray}\label{ThEvol}
    i\frac{d\psi_{n}}{dz} &= -J'_{n} \psi_{n+1} - J'_{n-1} \psi_{n-1} \nonumber \\ 
    J'_{n}\left(z\right)  &= \begin{cases}
                               J_{n}\left(z\right), & J_{n}\left(z\right) \geq J_{Th}, \\
                               0, & J_{n}\left(z\right) < J_{Th} \end{cases}.
\end{eqnarray} 

This intuitive mapping between the AAH model with real nonlinearity $g \left|\psi_{n}\right|^2$ (eq.\ref{NLEvol}) and the AAH model with the synthetic nonlinearity (eq.\ref{ThEvol}) is inspired by a system of two coupled nonlinear waveguides (Fig.\ref{Fig:1}e). The spatial evolution of the fields in this system is described using the equations
\begin{eqnarray} \label{TwoWG}
i\frac{d\psi_{1}}{dz} &= -J \psi_{2} - g \left|\psi_{1}\right|^2 \psi_{1} \nonumber\\
i\frac{d\psi_{2}}{dz} &= -J \psi_{1} - g \left|\psi_{2}\right|^2 \psi_{2},
\end{eqnarray} 
which are similar in form as eq.\ref{NLEvol} but with uniform coupling strength $J$. A numerical solution to these equations, using the fourth-order Runge-Kutta method, shows that when light is injected into one of the waveguides $\left(\psi_{1}\left(z = 0\right) = 1, \psi_{2}\left(z = 0\right) = 0 \right)$, the maximum power $P_{2} = \left|\psi_{2}\right|^{2}$ coupled to the second waveguide shows a sharp decrease with increasing nonlinear strength $g$. This indicates a breakdown of the effective coupling between the waveguides because of the nonlinearity-induced on-site potential difference or, equivalently, an effective refractive-index difference between the two waveguides. The greater the coupling (the smaller the gap) between the waveguides, the greater is the nonlinear strength required to break the coupling. Therefore, introducing $J_{Th}$ in eq.\ref{ThEvol} effectively simulates the role of nonlinearity in eq.\ref{NLEvol} such that a particular choice of $J_{Th}$ qualitatively corresponds to a particular choice of $g\left|\psi_{n}\right|^2$. 

To demonstrate soliton-like Thouless pumping using our approach, we consider the AAH model with periodicity $N = 3$, coupling strengths $J_{0} = 1, ~\delta = 0.8$, and period $\Omega = 1/\Lambda = 1/20$. A schematic of the coupled waveguide array that implements this model and the sinusoidal modulation of the coupling strengths is shown in Fig.\ref{Fig:1}a,b. As expected, this model exhibits three bulk bands, with Chern numbers $C = -1, 2, -1$ (Fig.\ref{Fig:1}c). The spatial evolution of an excitation wavefunction corresponding to the Wannier function of the second band ($C = 2$) shows diffraction such that its center of mass (CoM), defined as $\sum_{n} n\left|\psi_{n}\right|^{2}$, shows pumping by two unit cells (6 lattice sites) in a single period of the drive (Fig.\ref{Fig:1}d,k).

\begin{figure}[h]
    \centering
    \includegraphics[width=0.46\textwidth]{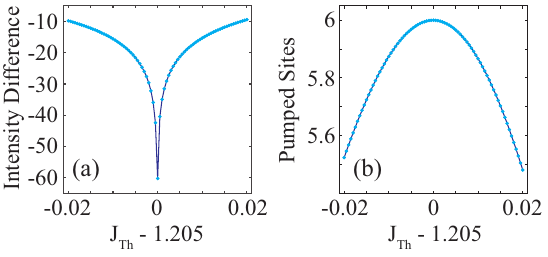}
    \caption{\textbf{a}, Variation of intensity difference between input and output soliton state, and \textbf{b}, the number of pumped lattice sites as a function of $J_{Th}$.}
    \label{Fig:3}
\end{figure}

\begin{figure*}[ht!]
    \centering
    \includegraphics[width=\textwidth]{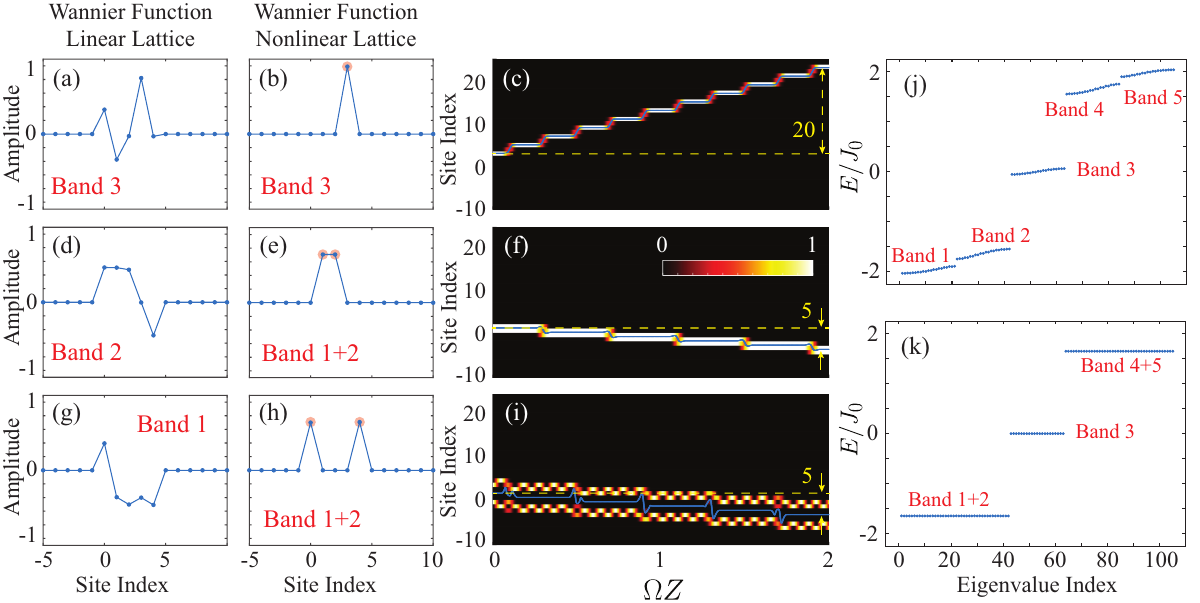}
    \caption{Wannier functions corresponding to Band 3, for \textbf{a}, linear lattice, and \textbf{b}, effectively nonlinear lattice. \textbf{c.} Quantized pumping for Band 3 by 2 unit cells per period. \textbf{d} Wannier function for Band 2. \textbf{e} Composite Wannier function for degenerate bands 1 and 2. \textbf{f.} Fractional pumping by one unit cell in two periods. \textbf{g}. Wannier function for Band 1 of the linear lattice. \textbf{h, i} Wannier function and fractional pumping for the second Wannier function of degenerate bands 1 and 2. \textbf{j,k} Energy eigenvalues at $z=0$ for the linear lattice and the nonlinear lattice showing degeneracy between bands 1 and 2.}
    \label{Fig:4}
\end{figure*}
The schematic of the effectively nonlinear coupled waveguide lattice with thresholded couplings and the resulting band structure is shown in Figs.\ref{Fig:1}g-i. We chose $\phi_{0} =  \frac{2\pi}{10}$ and first excite a single site within a single unit cell, $\left( \Psi_{in} = [1, 0, 0] \right)$. By tuning $J_{Th} \simeq 1.205 J_{0}$, we clearly observe soliton-like pumping where there is absolutely no diffraction, but the excited wavefunction is displaced by exactly two unit cells (6 lattice sites), corresponding to the Chern number of the second band (Fig.\ref{Fig:2}c). Interestingly, this excitation does not correspond to the Wannier function of the uniformly filled second band of the linear AAH lattice (Fig.\ref{Fig:2}a), which occupies all three lattice sites within a unit cell. Therefore, with this excitation, we do not expect to observe quantized pumping in the linear lattice. Nevertheless, we find that this single-waveguide excitation does indeed represent the Wannier function of the uniformly filled second band ($C = 2$) of the effectively nonlinear lattice, and this leads to the observation of quantized pumping (Fig.\ref{Fig:2}b). This observation completely aligns with the phenomenon of quantized Thouless pumping in linear lattices, which requires the excitation of a Wannier function for a uniformly filled band. Furthermore, in linear lattices, the evolution of CoM in the lattice tracks the instantaneous Wannier function of the band. We observe a similar behavior where the single-site soliton exactly corresponds to the single-site Wannier function of the effectively nonlinear lattice for the second band (Fig.\ref{Fig:2}d).

We observe similar quantized pumping for unit-cell excitations $\Psi_{in} = [0, 1/\sqrt{2}, 1/\sqrt{2}]$  and $\Psi_{in} = [0, 1/\sqrt{2}, -1/\sqrt{2}]$, which correspond to the Wannier functions of the first and third bands of the nonlinear lattice, respectively (Fig.\ref{Fig:2}e-l). Because both these bands exhibit $C = -1$, we observe displacement of the excitation wavepacket by one unit cell (3 sites), but in the opposite direction to that of the second band. Once again, these excitations do not correspond to the Wannier functions of the filled bands of the linear lattice, but they do correspond to those of the thresholded nonlinear lattice. We also observe that the pumped two-site solitons track the instantaneous Wannier functions of the corresponding bands (Figs.\ref{Fig:2}h,l).  

We note that setting $J_{Th} = J_{0} + \delta$, the maximum coupling strength, effectively breaks all couplings between the lattice sites. This condition corresponds to a very high nonlinear strength such that the wavepacket is self-localized or trapped in a single site and shows no pumping \cite{Jurgensen2021}.

We also note that the use of thresholded couplings creates a non-adiabatic evolution of the wavepacket (Fig.\ref{Fig:1}g-i). Nevertheless, topological pumping requires adiabatic evolution only to the extent that the spatial/temporal modulation does not generate any mixing between Wannier functions of bands with different Chern numbers. Indeed, non-adiabatic fast topological pumps have recently been demonstrated \cite{Fedorova2020, Song2024}. In our lattice, despite non-adiabaticity, we do not observe any such mixing, and as Fig.\ref{Fig:2} shows, the observed pumping is precisely dictated by the Chern number of the corresponding band. 

A hallmark feature of solitons is that they balance diffraction/dispersion against the spatial/temporal phase shifts introduced by nonlinearity and, therefore, preserve their shape and peak intensity during evolution. Consequently, both the formation of solitons and their quantized pumping are observed only for a narrow range of the nonlinear strength $g$ (with normalized, unit intensity) \cite{Jurgensen2021, Mostaan2022}. The solitons formed in our effectively nonlinear lattice exhibit an analogous sensitivity to the threshold coupling strength $J_{Th}$ (Fig.\ref{Fig:3}). Specifically, we analyzed the intensity difference between the input and the output soliton wavepacket and the number of pumped waveguides as $J_{Th}$ was varied. At an optimal value $J_{Th} \simeq 1.205$, the intensity difference between the input and output soliton is as small as $10^{-6}$ (-60 dB), limited only by the precision of our numerical simulations (Fig.\ref{Fig:3}a). At this optimal $J_{Th}$, the number of pumped unit cells (in this case, 2) is also exactly equal to the Chern number of the corresponding band (Fig.\ref{Fig:3}b). As we deviate away from the optimal $J_{Th}$, soliton pumping is still observed, but the intensity difference increases and the number of pumped lattice sites decreases.

To show fractionally quantized pumping in our effectively nonlinear system, we consider a five-band AAH model, with $N = \frac{5}{2}$, and coupling strengths of the linear system described as
\begin{equation}
 J_{n}\left(z\right) = J_{0} + \delta ~cos\left(\frac{2\pi}{5/2} n + 2\pi \Omega z + \frac{2\pi}{10} \right).
\end{equation}
The Chern numbers of the five bands are $C = 2, -3, 2, -3, 2$, respectively (Fig.\ref{Fig:4}). As before, we introduce a threshold in the coupling strengths, with an optimal $J_{Th} \simeq 1.008$. When we excite a single lattice site in the unit cell, with $\Psi_{in} = [0, 0, 0, 1, 0]$, we observe integer quantized pumping of a soliton by 2 unit cells (10 lattice sites) in one period (Fig.\ref{Fig:4}c). We find that this excitation does not represent the Wannier function of any uniformly filled band of the linear lattice, but it represents the Wannier function of the filled third band of the nonlinear lattice (Fig.\ref{Fig:4}a,b). Therefore, the single-site excitation is displaced by two unit cells, equal to the Chern number of the third band. 

In contrast, when we chose the input excitation in a unit cell to be $\Psi_{in} = [0, 1/\sqrt{2}, 1/\sqrt{2}, 0, 0]$, we observe fractionally quantized pumping of a soliton that is displaced only by one unit cell in two periods, similar to that observed in ref. \cite{Jurgensen2023}. This fractional quantization can be understood by considering the eigenvalues of the finite lattice at the input (Fig.\ref{Fig:4}j,k). Although the linear lattice does exhibit five nondegenerate bands, the introduction of effective nonlinearity introduces degeneracy between Bands 1 and 2, and Bands 4 and 5. Therefore, this input excitation corresponds to a multi-band Wannier function composed of Bands 1 and 2, with Chern numbers $2$ and $-3$, respectively, of the nonlinear lattice. Accordingly, the observed displacement $f = -1/2$ of one unit cell in two periods corresponds to $\sum_{i=1}^{N_{d}} C_{i}/N_{d}$, that is, the average of the Chern numbers $\left(C_{1} = 2, ~C_{2} = -3\right)$ of the $N_{d} = 2$ number of degenerate bands \cite{Jurgensen2023}. Note that, as before, this excitation does not uniformly fill any band or a pair of bands of the linear lattice. 

When we excite the second Wannier eigenstate of the composite bands 1 and 2, with $\Psi_{in} = [1/\sqrt{2}, 0, 0, 0, 1\sqrt{2}]$, we observe pumping of a soliton molecule, that is, a pair of solitons. Each of these solitons is also pumped by one unit cell in two periods, equal to the average Chern number of Bands 1 and 2. We observe similar fractionally quantized pumping, of one unit cell in two periods, by exciting the composite Wannier functions of Bands 4 and 5. 

\begin{figure}
    \centering
    \includegraphics[width=\columnwidth]{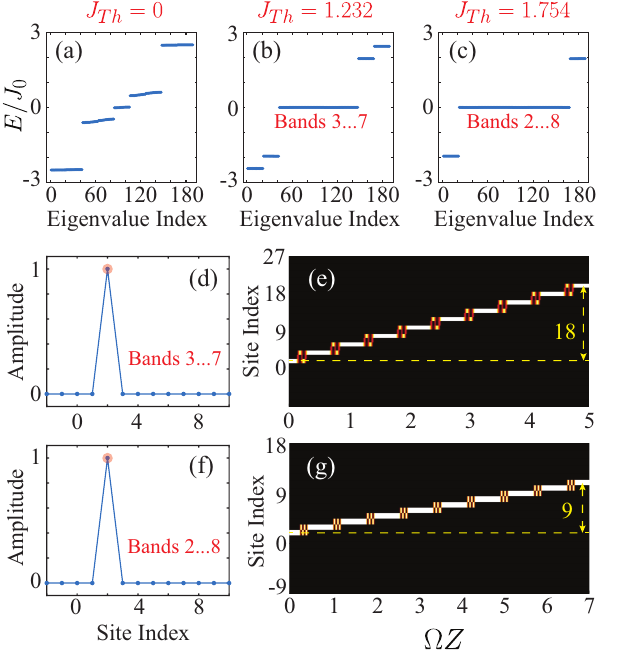}
    \caption{Eigenvalues for a 9-band AAH model, with PBC and 21 cells, for \textbf{a.} $J_{Th} = 0$, \textbf{b.} $J_{Th} \simeq 1.232$, \textbf{c.} $J_{Th} \simeq 1.754$. 5 bands and 7 bands are degenerate in \textbf{b} and \textbf{c}, respectively, at zero energy. \textbf{d.} One of the Wannier functions corresponding to the degenerate 5 bands of \textbf{b}, showing occupation of a single lattice site. \textbf{e.} Observed pumping of 2 unit cells in 5 periods. \textbf{f.} Wannier function for the 7 degenerate bands of \textbf{c}, with the same occupation as \textbf{d}. \textbf{g}. Pumping by 1 unit cell in 7 periods. 
    }
    \label{Fig:5}
\end{figure}

Fractionally quantized soliton pumping exhibits another intriguing feature, where the number of pumped unit cells in a period depends on the strength of nonlinearity $g$ \cite{Jurgensen2023,Fu2022}. To show similar behavior in our effectively nonlinear system, we consider a 9-band AAH model with $J_{0} = 1, ~\delta = 0.95$, and equivalent magnetic flux $\phi = 2\pi \frac{2}{9}$. The Chern numbers for these nine bands are $C = \left(4, -5, 4, -5, 4, -5, 4, -5, 4 \right)$. As expected, the eigenvalues for the linear system $\left(J_{Th} = 0 \right)$ with 21 unit cells and PBC show 9 bands (Fig.\ref{Fig:5}a). Introducing an effective nonlinearity, with $J_{Th} \simeq 1.232$, creates a degeneracy between five bands with their energy $E = 0$. This degeneracy at $E = 0$ is because of the introduction of $J_{Th}$, which sets the coupling strengths between some lattice sites to zero.

When we excite a single site of a unit cell, corresponding to one of the Wannier functions of these five degenerate bands, we find that the soliton wavpacket is pumped by exactly two unit cells in five periods. This displacement matches the expected pumping $f = \left(4 - 5 + 4 - 5 + 4\right)/5 = 2/5$. Nevertheless, on increasing $J_{Th} \simeq 1.754$, we find another regime where seven bands, with $C = \left(4, -5, 4, -5, 4, -5, 4\right)$ are degenerate at $E = 0$. Now, the excitation of a single site of the unit cell (same as in \textbf{d.}), also corresponding to one of the Wannier functions of the degenerate 7 bands, shows that the soliton is pumped by only one unit cell in 7 periods, with $f = 1/7$.

\begin{figure}
    \centering
    \includegraphics[width=\columnwidth]{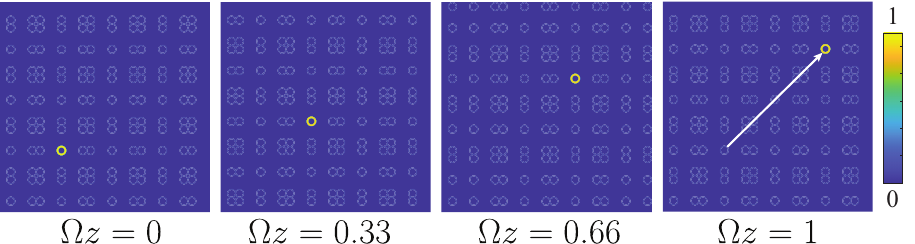}
    \caption{Quantized topological pumping in a 2D 3-band AAH model, showing quantized displacement by two unit cells (6 lattice sites) in both spatial directions, during one pump cycle. 
    }
    \label{Fig:6}
\end{figure}

Finally, we show that our approach of soliton-like pumping using effective
nonlinearities can also be generalized to 2D soliton pumping using a simple extension of 1D pumping in two independent directions \cite{Jurgensen2022}. In particular, we chose the 2D version of the 3-band AAH model (eq.\ref{NLEvol}), which corresponds to the 4D quantum Hall effect \cite{Zilberberg2018, Lohse2018}. Fig.\ref{Fig:6} shows the quantized pumping observed for the exciting of a single lattice site, corresponding to the Wannier function of Band 2 ($C = +2$) of the nonlinear lattice. As expected, the excitation is displaced by exactly two unit cells (6 lattice sites) in both directions. We note that we chose exactly the same $J_{Th}$ as that of Fig.\ref{Fig:2}. This simple extension to 2D systems clearly shows the versatility of our approach. 

To summarize, we have shown that linear AAH lattices with a synthetic nonlinearity, in the form of thresholded coupling strengths, exhibit quantized and fractionally quantized topological pumping similar to that observed in nonlinear AAH models. More importantly, our results reconcile Thouless pumping of nonlinear solitons with conventional linear Thouless pumping, showing that both require the excitation of a Wannier function of a uniformly filled band of the linear or the nonlinear lattice. Such soliton-like pumping could be experimentally implemented using synthetic lattices in time and frequency, which allow for the dynamical tuning of the threshold coupling strength \cite{Wimmer2015, Yuan2018, Chalabi2019, Sridhar2024}. Our approach could also be applied to reveal the interplay between nonlinear, non-Hermitian, and higher-order topological physics in other model systems where understanding the formation of solitons and their evolution remains challenging.
\\
\begin{acknowledgments}
This research was supported by the National Science Foundation (DMREF-2323908). 
\end{acknowledgments}


\bibliography{Soliton_Pumping_arXiv.bbl}

\end{document}